\documentclass[fp,twocolumn]{jpsj3}
\usepackage{txfonts}
\usepackage{color}

\title{Influence of the Fermi Surface Morphology on the Magnetic Field-Driven Vortex Lattice Structure Transitions in YBa$_{2}$Cu$_{3}$O$_{7-\delta}:\delta=$~0, 0.15}

\author{Nikola Galvan Leos$^1$, Jonathan S. White$^{1,2,}$\thanks{jonathan.white@psi.ch}, Joshua A. Lim$^{3,4,5}$, Jorge L. Gavilano$^1$, Bernard Delley$^6$, Louis Lemberger$^{3,7}$, Alexander T. Holmes$^3$, Marisa Medarde$^8$, Toshinao Loew$^9$, Vladimir Hinkov$^{10}$, Chengtian Lin$^{9}$, Mark Laver$^{11}$, Charles D. Dewhurst$^{7}$ and Edward M. Forgan$^3$}
\inst{$^1$Laboratory for Neutron Scattering and Imaging, Paul Scherrer Institut, CH-5232 Villigen, Switzerland \\
$^2$Laboratory for Quantum Magnetism, \'{E}cole Polytechnique F\'{e}d\'{e}rale de Lausanne (EPFL), CH-1015 Lausanne, Switzerland \\
$^3$School of Physics and Astronomy, University of Birmingham, Edgbaston, Birmingham, B15 2TT, UK\\
$^4$Institut f\"{u}r Festk\"{o}perphysik, TU Dresden, D-01069 Dresden, Germany\\
$^5$Heinz Maier-Leibnitz Zentrum (MLZ), Lichtenbergstr.\ 1, D-85748 Garching, Germany\\
$^6$Condensed Matter Theory Group, Paul Scherrer Institut, CH-5232 Villigen, Switzerland\\
$^7$Institut Laue-Langevin, 6 rue Jules Horowitz, 38042 Grenoble, France\\
$^8$Laboratory for Developments and Methods, Paul Scherrer Institut, CH-5232 Villigen, Switzerland\\
$^9$Max Planck Institut f\"ur Festk\"orperforschung, D-70569 Stuttgart, Germany\\
$^{10}$Physikalisches Institut und R\"ontgen Center for Complex Materials Systems, Universit\"at W\"urzburg, 97074 W\"urzburg, Germany\\
$^{11}$School of Metallurgy and Materials, University of Birmingham, Edgbaston, Birmingham, B15 2TT, UK} 

\abst{\color{black}We report small-angle neutron scattering measurements of the vortex lattice (VL) structure in single crystals of the lightly underdoped cuprate superconductor YBa$_2$Cu$_3$O$_{6.85}$. At 2~K, and for fields of up to 16~T applied parallel to the crystal \textbf{c}-axis, we observe a sequence of field-driven and first-order transitions between different VL structures. By rotating the field away from the \textbf{c}-axis, we observe each structure transition to shift to either higher or lower field dependent on whether the field is rotated towards the [100] or [010] direction. We use this latter observation to argue that the Fermi surface morphology must play a key role in the mechanisms that drive the VL structure transitions. Furthermore, we show this interpretation is compatible with analogous results obtained previously on lightly overdoped YBa$_2$Cu$_3$O$_{7}$. In that material, it has long-been suggested that the high field VL structure transition is driven by the nodal gap anisotropy. In contrast, the results and discussion presented here bring into question the role, if any, of a nodal gap anisotropy on the VL structure transitions in both YBa$_2$Cu$_3$O$_{6.85}$ and YBa$_2$Cu$_3$O$_{7}$.}

\begin{document}
\maketitle

\section{Introduction}
\label{sec:1Int}
It is now more than fifty years since the seminal theoretical work of Abrikosov wherein the superconducting vortex lattice (VL) was first shown to be expected in type-II superconductors.~\cite{Abr57} This pioneering work sparked the intense and continuing research efforts into the physics of vortex matter, which assumed even greater importance with the discoveries of High-$T_{\rm c}$ cuprate and other unconventional type-II materials with novel pairing symmetries. A full understanding of the properties of vortex matter is also crucial from the technological perspective because type-II superconducting materials are, for example, key constituents of high current-carrying applications. From the latter viewpoint, it becomes vital to understand the vortex properties of High-$T_{\rm c}$ cuprates that display superconducting critical temperatures above those of the 77~K boiling point of liquid nitrogen, since this is a much more economical cryogen than liquid helium.

Close to optimally-doped YBa$_2$Cu$_3$O$_{7-\delta}$ is a prototypical High-$T_{\rm c}$ cuprate with superconducting critical temperatures in the region of 90~K, the exact value being composition dependent.~\cite{Lia06} Indeed, the VL properties of YBa$_2$Cu$_3$O$_{7-\delta}$ have been the subject of numerous investigations over the years,~\cite{Gam87,Mag95,Sch96,Fis97,Son97,Son99,Aus08,Kie10,For90,Yet93a,Yet93b,Kei93,Kei94,For95,Joh99,Bro04,Sim04,Whi08,Whi09,Cam14} with the bulk probe of small-angle neutron scattering (SANS) playing an important role in revealing the effects of effective mass anisotropy, VL melting and pinning, and field-induced nonlocality.~\cite{For90,Yet93a,Yet93b,Joh99,Bro04,Kei93,Kei94,For95,Sim04,Whi08,Whi09,Cam14} Despite the large body of work, it is only in the last decade that SANS observations of the intrinsic magnetic field- and temperature-dependent VL structures could be achieved.~\cite{Whi09,Whi11,Cam14} The more recent achievements can be largely attributed to significant improvements in sample quality, since these observations were not complicated by twin-plane pinning effects which hampered the earlier efforts to unveil the intrinsic VL properties.~\cite{For90,Yet93a,Yet93b,Joh99,Bro04}

In particular, at low temperature in detwinned and lightly overdoped YBa$_2$Cu$_3$O$_{7}$, the field-evolution of the VL structure was recently studied for magnetic fields up to 16.7~T applied parallel to the crystal \textbf{c}-axis.~\cite{Whi09,Whi11,Cam14} The most striking observations include successive field-driven first-order reorientation transitions of the VL coordination at both 2.3(2)~T and 6.7(2)~T,~\cite{Whi09} and a smooth evolution of the high-field VL that passes \emph{through} the perfectly square VL coordination expected to be stabilised by a pure $d$-wave gap symmetry.~\cite{Cam14} In general, anisotropies in both the Fermi surface and the superconducting gap can be responsible for the field-dependence of the precise VL structure,~\cite{Aff97,Nak02,Suz10} and also cause VL structure transitions. However, and as discussed in Refs.~\cite{Whi11,Cam14}, YBa$_2$Cu$_3$O$_{7}$ remains a high-profile example of an unconventional $d$-wave superconductor in which the physical origins of the VL structure transitions are not completely understood.

Towards addressing this problem, here we present new SANS studies of detwinned and lightly underdoped YBa$_2$Cu$_3$O$_{6.85}$. Despite the light underdoping which introduces oxygen vacancies, and hence pinning centers, into the CuO chains, we nonetheless observe well-ordered VLs up to the highest magnetic field of 16~T applied parallel to the \textbf{c}-axis. We observe a sequence of VL structure transitions in this material that is analogous to that seen in YBa$_2$Cu$_3$O$_{7}$, though the transition fields are slightly modified. We also present extensive measurements of the VLs observed in YBa$_2$Cu$_3$O$_{6.85}$ as a function of angle of applied field to the \textbf{c}-axis. The VL structure transition fields are found to display a strong and characteristic angle-dependence, and we use this observation to provide new insights into the physical origin of the two field-driven VL structure transitions in YBa$_2$Cu$_3$O$_{7-\delta}$.

\section{Experimental Method}
\label{sec:2Exp}
Single crystals of YBa$_2$Cu$_3$O$_{6.85}$ were prepared by the flux growth method, and from a starting mixture of BaCO$_3$, CuO and Y$_2$O$_3$ powders molten in yttrium-stabilized ZrO$_{2}$ crucibles.~\cite{Lin91} The as-grown crystals are crystallographically twinned and also have an ill-defined oxygen content. The individual crystals were therefore detwinned at elevated temperature using a uniaxial stress applied along a $\langle 100\rangle$ axis. To obtain the YBa$_2$Cu$_3$O$_{6.85}$ composition, the crystals were subsequently annealed at 520$^{\circ}$~C in an O$_{2}$ atmosphere.~\cite{Hin07}

The sample used for the SANS measurements was a co-aligned mosaic of eight YBa$_2$Cu$_3$O$_{6.85}$ single crystals of total mass 85~mg. Using single crystal neutron diffraction to study the (200) and (020) nuclear reflections, the bulk twin-domain population ratio was determined for each individual crystal. The YBa$_2$Cu$_3$O$_{6.85}$ single crystals were found to be almost completely twin-free, and the entire mosaic displayed a detwinning ratio of $98.6(5)\%$. The superconducting critical temperature ($T_{\rm c}$) of each individual crystal was also measured using a Quantum Design physical property measurement system (PPMS). The overall sample displayed a $T_{\rm{c}}$ of 90.5(1.0)~K, where the error represents the spread in $T_{\rm{c}}$ amongst the crystals. Comparison between this $T_{\rm{c}}$ value and the data reported in Ref.~\cite{Lia06} gives a sample doping level of $\delta=0.15(1)$. The sample mosaic was co-aligned and mounted on a 1~mm thick Si plate with the crystal \textbf{c}-axis of the sample perpendicular to the plate.

The SANS experiments were conducted on the SANS-I instrument at the Swiss spallation neutron source (SINQ), Paul Scherrer Institut (PSI), Villigen, Switzerland, the D11 at the Institut Laue-Langevin (ILL), Grenoble, France, and the NG3-SANS instrument at NIST Center for Neutron Research, USA. The neutrons were collimated over 6-18~m (SANS-I), 5.5~m (D11) and 4.5-8m (NG3-SANS), and wavelengths between 4.5-10~\AA~were selected with a typical FWHM spread of $10\%$. A position-sensitive 2D multi-detector placed behind the sample was used to detect the scattered neutrons. At SANS-I, D11, and NG3-SANS, 11~T, 17~T and 9~T horizontal field SANS cryomagnets were respectively available. In each cryomagnet the sample was initially installed with the crystal \textbf{c}-axis parallel to the applied field direction, and approximately parallel to the neutron beam. For the 11~T magnet at PSI, an additional motor-head allowed rotation of the sample stick about the vertical axis inside the variable temperature insert (VTI) of the cryomagnet. This motor provided in-situ control of the direction of the magnetic field with respect to the \textbf{c}-axis, therefore enabling the systematic field-angle-dependent measurements of the VL structure presented in Section~\ref{sec:3Res_VL_Ang}.

At each SANS instrument the cryomagnet was placed on a goniometer table that enabled both rotation and tilt of the sample-magnet ensemble. The SANS measurements of the VL were done by rotating the sample and cryomagnet together through a range of angles that moved the VL diffraction peaks through the Bragg condition. By summing over all measurements at different angles, the SANS diffraction patterns shown in this paper are obtained, and wherein all the VL diffraction spots can be observed in a single image.

For all the low temperature measurements reported in this paper, the VL was prepared using the oscillation-field-cooling (OFC) method. By this method the sample is field-cooled through $T_{\rm{c}}$ until the measurement temperature in a time-dependent magnetic field that weakly oscillates around the target field. The oscillation amplitude used was typically 1~\% of the target field. The OFC method was used to prepare the VL since previous work on twin-free YBa$_2$Cu$_3$O$_{7}$ showed such an approach allows i) the VL to achieve an improved perfection, and ii) a preferred VL coordination due to a lower effective temperature than achieved when cooling in a static field.~\cite{Whi09} Additional SANS measurements not reported here confirmed that the OFC method produced a similar effect on the VLs observed in the present YBa$_2$Cu$_3$O$_{6.85}$ sample. Background SANS measurements were conducted at $T=95$~K and subtracted from the low temperature measurements in order to leave just the VL signal. SANS data visualization and analysis was carried out using the GRASP software.~\cite{Dew03}

\section{Results}
\label{sec:3Res}

\subsection{Vortex lattice structure for $\mu_{0}H\parallel$~c in YBa$_2$Cu$_3$O$_{6.85}$}
\label{sec:3Res_VL_Coord}
\begin{figure}
\centering
\includegraphics[width=0.48\textwidth]{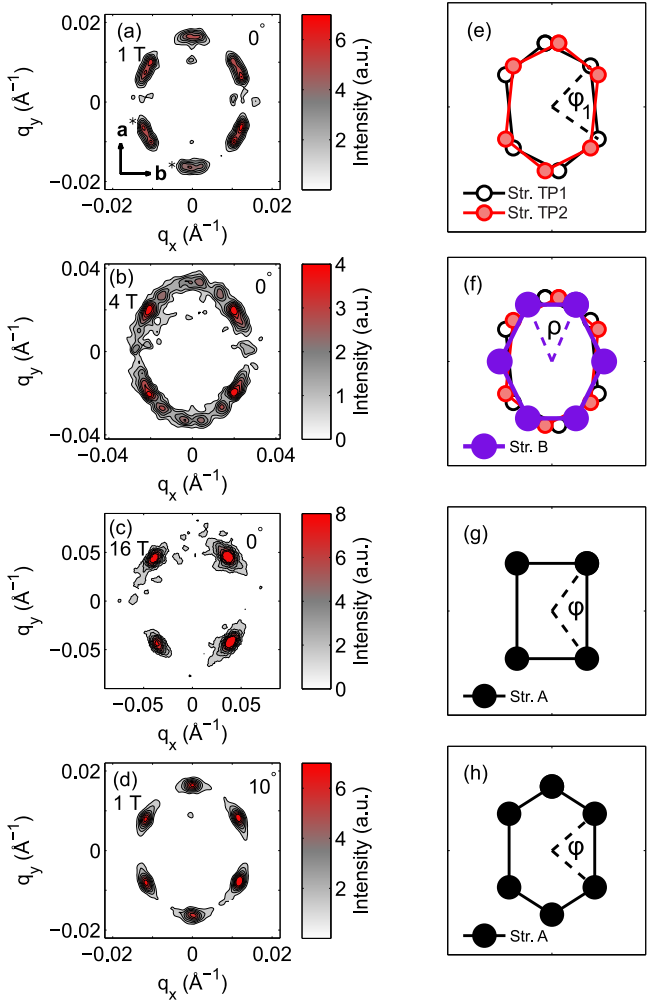}
\caption{(Color online) SANS diffraction patterns obtained from the VL at 2~K and (a) $\mu_{0}H\parallel$~\textbf{c}$~=1$~T, (b) $\mu_{0}H\parallel$~\textbf{c}$~=4$~T, (c) $\mu_{0}H\parallel$~\textbf{c}$~=16$~T, and (d) $\mu_{0}H$~=1~T applied $10^{\circ}$ away from \textbf{c}. Panels (e) to (h) show schematic drawings of the results shown in (a) to (d), and which clarify the different VL domains present. Panels (e) and (f) each show VL domains, denoted TP1 and TP2, whose alignment is affected by pinning effects due to the small residual density of twin-planes. In contrast, structures B (panel (f)) and A (panels (g) and (h)) are aligned with the crystal axes and their alignment is considered to be intrinsic.}
\label{Fig:1}
\end{figure}

Figs.~\ref{Fig:1}(a)-(c) show a selection of VL SANS diffraction patterns obtained from YBa$_2$Cu$_3$O$_{6.85}$ at $T=$~2~K, and at different magnetic fields applied parallel to the crystal \textbf{c}-axis ($\mu_{0}H\parallel$~\textbf{c}). Fig.~\ref{Fig:1}(a) shows the diffraction pattern obtained at 1~T that is typical for fields up to 2.37(2)~T. As shown schematically in Fig.~\ref{Fig:1}(e), this VL is composed of two distorted hexagonal domains that each have two spots aligned with a $\{110\}$ plane. The alignment of VL domains along these directions is a clear indication that the VL orientation is not intrinsic, but instead controlled by the small volume of twin-planes remaining in the sample.~\cite{Bro04,Sim04,Whi08} Consequently, we refer to this two domain VL structure as structure TP. As illustrated for one of the two domains in Fig.~\ref{Fig:1}(e), the primitive cell opening angle $\phi_{1}$ is almost bisected by the horizontal \textbf{b}$^{\ast}$-axis, but not exactly.

In the field range between 2.37(2)~T and 5.37(2)~T, structure TP coexists with an additional single distorted hexagonal VL domain, which we call structure B. Fig.~\ref{Fig:1}(b) shows the typical SANS pattern for this field range, with a schematic illustration of all the VL domains present shown in Fig.~\ref{Fig:1}(f). In contrast to structure TP, structure B is aligned intrinsically with the crystal axes, and its primitive unit cell is bisected exactly by the vertical \textbf{a}$^{\ast}$-axis. Across this field range the VL domains of both structures TP and B are always observed to co-exist and neither exclusively populates the entire sample. By surveying the individual crystals of the sample mosaic, we confirmed the structure co-existence in all crystals, though the relative populations of each structure differed slightly. This observed variation could reflect small differences in either oxygen stoichiometry between the individual crystals, or a distribution in spacings between the residual twin planes.

Increasing the field beyond 5.37(2)~T leads to a suppression of both structures B and TP. Instead the VL structure becomes composed of a different intrinsic and single VL domain aligned with the crystal axes that we call structure A. Structure A prevails until the highest measured field of 16~T, the SANS diffraction pattern of which is shown in Fig.~\ref{Fig:1}(c). This structure is sketched in Fig.~\ref{Fig:1}(g), and we define the primitive unit cell to be bisected exactly by the horizontal \textbf{b}$^{\ast}$-axis. A general feature of the data obtained in the high-field range is that the high-field diffraction pattern is dominated by a rhombic VL coordination characterized by four intense Bragg spots, in contrast to the hexagonal VL domains observed at lower fields.

\begin{figure}[t]
\centering
\includegraphics[width=0.46\textwidth]{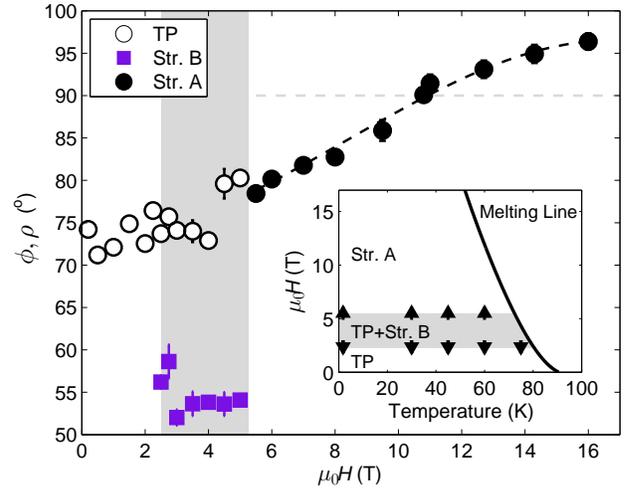}
\caption{(Color online) The VL opening angles of the various VL primitive cells for $\mu_{0}H\parallel$~\textbf{c} and at $T=$~2~K. The dashed line is a guide for the eye. The shaded region denotes the field-range of co-existence for structures TP and B. Inset: the $\mu_{0}H\parallel$~\textbf{c}-$T$ VL structure phase diagram. The symbols denote the transition fields between the different VL structure phases, as determined by tracking the VL structures in the vicinities of the phase boundaries. The melting line was obtained from Ref.~\cite{Ram12}.}
\label{Fig:2}
\end{figure}

Fig.~\ref{Fig:2} shows the $\mu_{0}H$-dependence at $T=2$~K of the primitive cell opening angles for structures TP, B, and A. For the TP structure, $\phi$ is an average taken over the two constituent domains. Overall, with increasing $\mu_{0}H$ there is a smooth evolution of the opening angle of the TP structure, before a first-order re-orientation transition which occurs at 5.37(2)~T into the high-field phase where only structure A exists. At higher field, the primitive cell of structure A evolves smoothly with increasing $\mu_{0}H$ up to 16~T, and at 10.6(2)~T it passes through the value of 90$^{\circ}$ expected for a perfectly square coordination. The co-existence of structures TP and B between 2.37(2)~T to 5.37(2)~T is also seen clearly, while for fields outside this field range just a single VL structure type populates the sample.

Bearing in mind the similarity between the primitive cell orientations for both structures TP and A, structure A can be expected to be the \emph{intrinsic} structure that would be present at lower fields in the absence of an orientational effect due to twin-plane pinning. This assumption can be readily tested, since it is well-known that in lightly-twinned YBa$_{2}$Cu$_{3}$O$_{7-\delta}$ samples, twin-plane pinning effects can be suppressed by rotating the sample so that all twin-planes form a small angle of $\sim$5-10$^{\circ}$ with respect to the field direction.~\cite{Bro04,Whi08} After rotating the present YBa$_{2}$Cu$_{3}$O$_{6.85}$ sample by 10$^{\circ}$ around the vertical \textbf{a}$^{\ast}$-axis with respect to a 1~T field, we find that such a rotation is indeed sufficient for the VL to overcome the twin-plane pinning. This is seen in Fig.~\ref{Fig:1}(d) which shows that in the rotated condition the VL diffraction pattern is composed of just a single distorted hexagonal domain aligned with the crystal axes, and not the $\{110\}$ planes. As illustrated in Fig.~\ref{Fig:1}(h), the primitive cell for this VL structure can be treated in the same way as structure A, with the denoted primitive unit cell bisected by the horizontal axis.

Returning to $\mu_{0}H\parallel$~\textbf{c}, another important feature seen in Fig.~\ref{Fig:2} is the broad field co-existence range of the TP and B structures. Such a co-existence exemplifies the near degeneracy between the two structure-types, and also shows that at 2.37(2)~T a first-order-type of VL structure transition occurs between a \emph{fraction} of the TP structure and structure B. Since such a `partial' transition occurs within \emph{all} crystals of the sample mosaic, it could be argued that the particular oxygen doping displayed by the sample appears to be close to a critical level for the occurrence of a complete phase transition. We also point out that this coexistence phase appears not to be a property of twin-plane pinning. As will be shown later, by again rotating the sample with respect to the field direction to suppress the pinning, the broad co-existence region survives, and is between structures A and B instead of structures TP and B.

The inset of Fig.~\ref{Fig:2} shows the $\mu_{0}H\parallel$~\textbf{c}-temperature VL structure phase diagram deduced from measurements where we tracked the VL transition fields. Here it is seen that the VL structure phase boundaries are only very weakly $T$-dependent, and so the field-width of the co-existence region remains essentially unchanged until the melting temperature. This behavior is similar to that reported for YBa$_{2}$Cu$_{3}$O$_{7}$ in Ref.~\cite{Whi11} with the same discussion presented there applicable also here for YBa$_{2}$Cu$_{3}$O$_{6.85}$.

\subsection{Vortex lattice structure for $\mu_{0}H$ not parallel to c in YBa$_2$Cu$_3$O$_{6.85}$}
\label{sec:3Res_VL_Ang}

\begin{figure}
\centering
\includegraphics[width=0.48\textwidth]{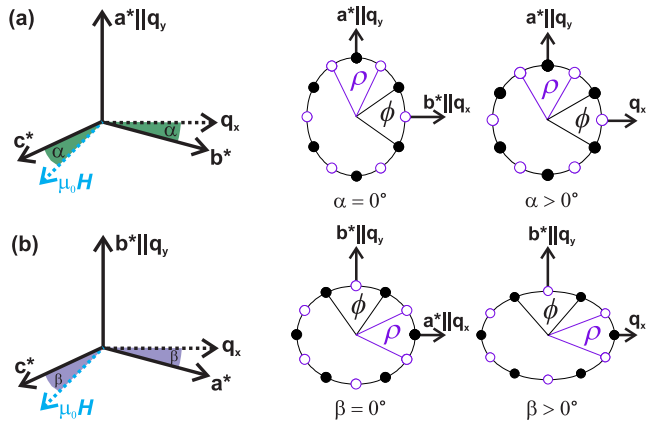}
\caption{(Color online) Sketches to illustrate the expected effect on the observed VL structures for SANS measurements done with the field applied at an angle to the crystal \textbf{c}-axis. In (a) the sample is rotated relative to the $\mu_{0}H$ axis by an angle $\alpha$ around \textbf{a}$^{\ast}$. In (b) the sample is rotated by an angle $\beta$ around \textbf{b}$^{\ast}$. On the right side of each panel are sketches of the expected SANS VL diffraction patterns for the different $\alpha$/$\beta$, and where directions $q_{x}$ and $q_{y}$ respectively denote the horizontal and vertical directions on the detector. The filled circles denote the Bragg peaks due to structure A, while empty circles those of structure B.}
\label{Fig:3}
\end{figure}

\begin{figure}
\includegraphics[width=0.3\textwidth]{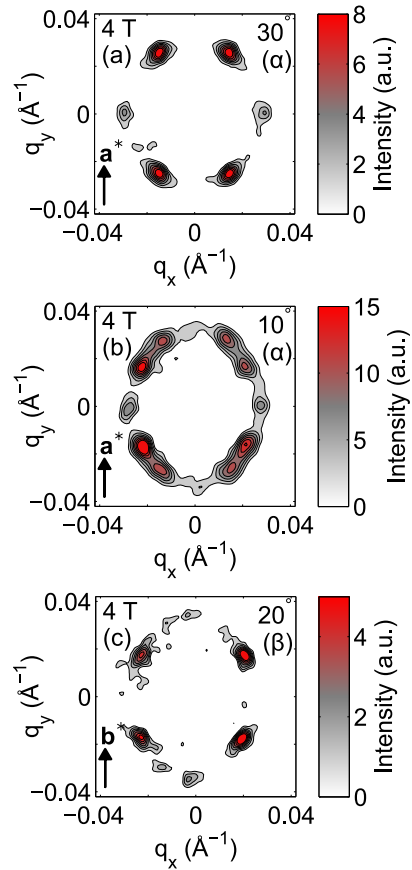}
\caption{(Color online) The VL structures observed at $T=$~2~K, and for $\mu_{0}H=$~4~T for (a) $\alpha=30^{\circ}$, (b) $\alpha=10^{\circ}$ and (c) $\beta=20^{\circ}$.}
\label{Fig:7}
\end{figure}

To provide further insight into the $\mu_{0}H$-dependent behaviour of the VL observed for $\mu_{0}H\parallel$~\textbf{c}, we conducted measurements of the VL structure for a range of angles between the applied field and the crystal \textbf{c}-axis. As illustrated in Fig.~\ref{Fig:3}, we explored two directions of sample rotation relative to $\mu_{0}H$: a) rotating the field by an angle $\alpha$ around the \textbf{a}$^{\ast}$-axis (towards $[010]$), and b) a rotation by an angle $\beta$ around the \textbf{b}$^{\ast}$-axis (towards $[100]$).

In Fig.~\ref{Fig:3} we also illustrate the change in distortion of hexagonal VL structures expected from anisotropic London theory as a consequence of the sample rotation. For $\mu_{0}H\parallel$~\textbf{c}, i.e. $\alpha=\beta=0^{\circ}$ we have seen in Sec.~\ref{sec:3Res_VL_Coord} that, although the orientation is determined by pinning, the low field hexagonal VLs are distorted along the \textbf{a}$^{\ast}$-axis away from an isotropic hexagonal coordination. Anisotropic London theory provides the simplest interpretation of the distortion in terms of an effective mass anisotropy since, $\eta=\gamma_{ab}=\lambda_{a}/\lambda_{b}\propto\sqrt{m_{a}^{\ast}/m_{b}^{\ast}}$,~\cite{Kog81,Cam88,Thi89} where $\eta$ is the axial ratio of the ellipse that overlays the six Bragg spots in a hexagonal VL domain. Since $\eta$ is related geometrically to the VL opening angles defined in Fig.~1, it is readily seen that $\eta>1$ at low field where London theory is most valid. In the zero-field limit we find $\eta=1.25(3)$ for YBa$_{2}$Cu$_{3}$O$_{6.85}$, which indicates a lower effective mass, and hence higher super-current density, along the \textbf{b}$^{\ast}$-axis compared with the \textbf{a}$^{\ast}$-axis; an observation consistent with the expected existence of superconducting CuO chain states.~\cite{Atk95,Xia96,Atk99} This suggestion is further supported by noting that the value of $\eta=1.25(3)$ for YBa$_{2}$Cu$_{3}$O$_{6.85}$, which has oxygen vacancies in the CuO chains, is less than that of $\eta=1.29(2)$ found for YBa$_{2}$Cu$_{3}$O$_{7}$ which has completely filled CuO chains,~\cite{Whi09,Whi11} and significantly less than the value of $\eta=2.57(5)$ reported for the stoichiometric double-chain compound YBa$_{2}$Cu$_{4}$O$_{8}$.~\cite{Whi14}

By increasing either $\alpha$ or $\beta$ the supercurrent response in the plane orthogonal to the field includes a contribution due to $m_{c}^{\ast}$. Since $m_{c}^{\ast}$$\gg m_{a}^{\ast}$ and $m_{b}^{\ast}$, increasing either $\alpha$ or $\beta$ always leads to an increase of the effective mass in the horizontal direction. Therefore, and as shown in Fig.~\ref{Fig:3}, increasing $\alpha$ causes the hexagonal VL distortion to decrease, and so a reduction of angle $\phi$ characteristic of structure A, or an \emph{increase} in angle $\rho$ characteristic of structure B. The converse is realised for increasing $\beta$ since the VL distortion will increase, with $\phi$ increasing and $\rho$ decreasing. It should be noted that even while anisotropic London theory provides a handy guide for understanding the angle-dependence of the VL distortion, no field-driven VL structure \emph{transitions} are expected within the theory for fields applied away from perpendicular to the basal plane. Therefore, any observed VL structure transitions still can only be understood as a consequence of higher-order anisotropies beyond those included in local anisotropic London theory. In what follows, it is observed that the $\mu_{0}H$-dependence of VL structure transitions are indeed strongly dependent on the direction of crystal rotation with respect to the field.

The usefulness of VL measurements with $\mu_{0}H$ at an angle to the crystal \textbf{c}-axis is exemplified by the data shown in Fig.~\ref{Fig:7}. Here we present VL diffraction patterns all obtained in $\mu_{0}H$=4~T but for different values of $\alpha$ and $\beta$. Beginning at $\mu_{0}H\parallel$~\textbf{c}, Fig.~\ref{Fig:1}(b) showed the VL at 4~T to be composed of structures TP and B. In contrast, with $\alpha=10^{\circ}$, an angle where pinning to twin-planes is suppressed, we observe the coexistence of structures A and B [Fig.~\ref{Fig:7}(b)]. Further increasing $\alpha$ to $30^{\circ}$ [Fig.~\ref{Fig:7}(a)] leads to a suppression of structure A, and structure B becomes the only VL structure type in the sample. On the $\beta$ side the inverse tendency with angle is observed. Fig.~\ref{Fig:7}(c) shows that for $\beta=20^{\circ}$, structure B is entirely suppressed, and structure A is the only structure-type in the sample. Therefore, at a constant field of 4~T we observe an angle-driven first-order VL structure transition between structures A and B. From the most general perspective, this observation is in agreement with the expectations of the `hairy-ball' theorem as applied to VLs, and which states that at any field discontinuous VL transitions must exist as the angle of the field to the crystal is varied.~\cite{Lav10,Kaw13}

\begin{figure}
\centering
\includegraphics[width=0.48\textwidth]{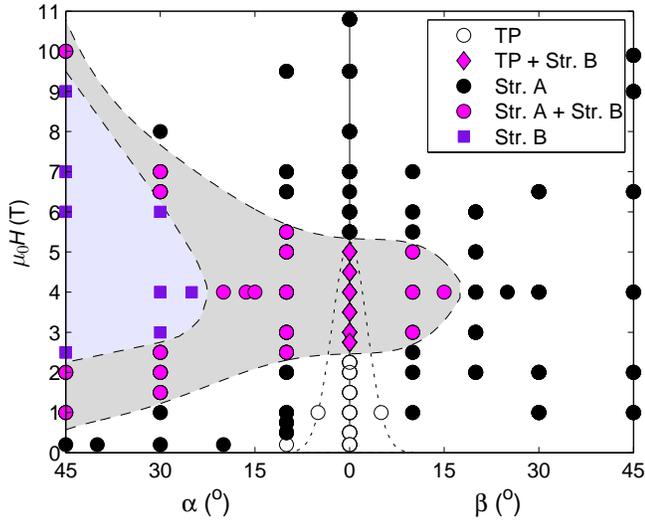}
\caption{(Color online) Magnetic field-$\alpha$/$\beta$ VL structure phase diagram at $T=2$~K. Each symbol denotes where the VL structure was measured by SANS. The phase boundary lines are estimated from the data. The dotted line close to zero angle ($\mu_{0}H\parallel$~\textbf{c}) and for fields below 5~T indicates the approximate region where vortex pinning can determine the VL orientation.}
\label{Fig:8}
\end{figure}

In Fig.~\ref{Fig:8} we show an overall angle-$\mu_{0}H$ VL structure phase diagram for YBa$_2$Cu$_3$O$_{6.85}$ that extends out to both $\alpha$ and $\beta=45^{\circ}$, and 10~T in applied field. Strongly contrasting $\mu_{0}H$-dependent behavior is observed for the two explored rotation directions. In general increasing $\alpha$ enhances the field stability range of structure B. Furthermore, for sufficiently large $\alpha$ the $\mu_{0}H$-dependent behavior evidences two \emph{complete} first-order transitions that separate structure B from structure A at both lower and higher field. Increasing $\beta$ leads to very different behavior, since any sign of structure B is quickly suppressed, and completely gives way to structure A at all fields by $\beta\geq17.5^{\circ}$.

\begin{figure}
\centering
\includegraphics[width=0.48\textwidth]{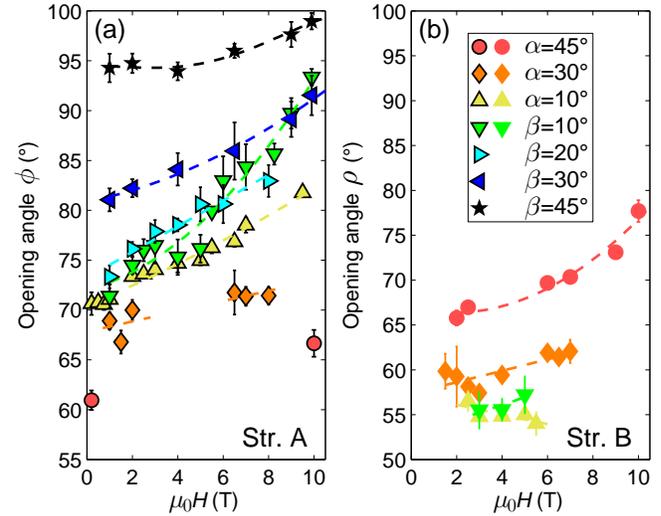}
\caption{(Color online) The $\mu_{0}H$-dependence of the VL opening angle of (a) structure A and (b) structure B over a range of $\alpha$ and $\beta$ rotation angles. All data were obtained at $T$=2~K, and all dashed lines are guides for the eye. The legend shown in (b) applies to both panels.}
\label{Fig:13}
\end{figure}

The phase diagram shown in Fig.~\ref{Fig:8} was constructed by tracking the $\mu_{0}H$-dependence of the opening angles of both structures A and B for the range of explored $\alpha$ and $\beta$ angles. These data are presented in Fig.~\ref{Fig:13}. The overall observed behavior is generally consistent with the expectation of anisotropic London theory illustrated in Fig.~\ref{Fig:3}. For structure A [Fig.~\ref{Fig:13}(a)], by increasing $\beta$, the value of $\phi$ at a particular field increases, which is consistent with an increase of the effective superconducting anisotropy in the plane orthogonal to the applied field. At low field, increasing $\alpha$ leads to a reduction in $\phi$ as the VL tends towards a more isotropic distortion. Fig.~\ref{Fig:13}(b) shows that angle $\rho$ of structure B also generally obeys the expected behavior as a consequence of the sample rotation.

Interpreting these angle-dependent measurements in terms of effective mass anisotropy incorporated into local London theory provides at least a qualitative understanding of why the VL opening angle, and hence distortion, changes as a consequence of rotating the crystal with respect to $\mu_{0}H$. Nonetheless, similarly as for $\mu_{0}H\parallel$~\textbf{c}, the most important aspects of the observations, the $\mu_{0}H$-dependence of both $\phi$ and $\rho$, and the transitions observed between structures A and B, require a theoretical explanation that goes beyond the local approximation inherent to the London theory.

\section{Discussion}
\label{sec:4Discussion}
The purpose of this study is to shed new light on the sequence of $\mu_{0}H$-driven and first-order VL structure transitions observed at low temperature in close to optimally-doped YBa$_{2}$Cu$_{3}$O$_{7-\delta}$ with $\mu_{0}H\parallel$~\textbf{c}. Using SANS we have studied the VL in a detwinned and lightly underdoped sample of YBa$_{2}$Cu$_{3}$O$_{6.85}$. By underdoping we aimed to controllably alter the electronic structure compared with the case of lightly overdoped YBa$_{2}$Cu$_{3}$O$_{7}$, both in terms of the precise Fermi surface morphology and the superconducting density of states (DOS). In theory, tuning these properties can lead to a variation in the $\mu_{0}H$-dependence of the VL structure between the two compounds.

First we establish the experimental observations. For $\mu_{0}H\parallel$~\textbf{c} and at $T$=2~K the following sequence of VL structure transitions has been reported in detwinned YBa$_{2}$Cu$_{3}$O$_{7}$:~\cite{Whi09,Whi11}
\begin{equation}
\mu_{0}H_{\rm{c1}} < \textrm{A} < 2.3(2)~\textrm{T} < \textrm{B} < 6.7(2)~\textrm{T} < \textrm{A} \nonumber
\end{equation}
where A/B denote VL structure types defined in the same way as in Sec.~\ref{sec:3Res}. This $\mu_{0}H$-dependent behavior is analogous to that reported here for YBa$_{2}$Cu$_{3}$O$_{6.85}$ which for $\mu_{0}H\parallel$~\textbf{c} and $T$=2~K is:
\begin{equation}
\mu_{0}H_{\rm{c1}} < \textrm{TP} < 2.37(2)~\textrm{T} < \textrm{TP}+\textrm{B} < 5.37(2)~\textrm{T} < \textrm{A}. \nonumber
\end{equation}
The principal differences between the two sequences concern the low and intermediate field phases. In the lowest field phase of YBa$_{2}$Cu$_{3}$O$_{6.85}$, measurements with $\mu_{0}H$ applied at an angle to the \textbf{c}-axis, allow us to deduce that structure TP arises due to residual twin-plane pinning in the sample. In a truly twin-free sample, structure A is expected to be present instead of structure TP, with this being the same low field VL structure-type seen in YBa$_{2}$Cu$_{3}$O$_{7}$.

In the intermediate field phase, both YBa$_{2}$Cu$_{3}$O$_{6.85}$ and YBa$_{2}$Cu$_{3}$O$_{7}$ display structure B, yet only in the latter compound does structure B occupy the entire sample for $\mu_{0}H\parallel$~\textbf{c}. In YBa$_{2}$Cu$_{3}$O$_{6.85}$, it remains unclear precisely why structures B and TP, which are spatially-separated in the sample, display such a broad range of field-coexistence. Since measurements at finite $\alpha$/$\beta$ show that the co-existence of structures TP and B becomes replaced by a co-existence between structures A and B, this allows us to rule out an obvious influence of the twin-plane pinning. Instead, the co-existence may reflect an extreme sensitivity of the VL structure to local variations in stoichiometry and homogeneity within the individual crystals of the sample. Whatever the true explanation, the general observation is that in terms of absolute field, the propensity for structure B falls with underdoping, since it occupies a narrower field range in YBa$_{2}$Cu$_{3}$O$_{6.85}$ compared with YBa$_{2}$Cu$_{3}$O$_{7}$. Future measurements on further underdoped samples may reveal a complete suppression of structure B for all $\mu_{0}H\parallel$~\textbf{c}.

It is also informative to re-write the transition sequences in terms of magnetic fields normalised to values of $H_{c2}(T=0)$, quantitative estimates of which have recently become available.~\cite{Gri14} For YBa$_{2}$Cu$_{3}$O$_{7}$, $H_{c2}$ is estimated to be 150(20)~T, and the transition sequence written in terms of normalized field is:
\begin{equation}
\mu_{0}H_{\rm{c1}} < \textrm{A} < 0.015(2) < \textrm{B} < 0.044(6) < \textrm{A} \nonumber
\end{equation}
For YBa$_{2}$Cu$_{3}$O$_{6.85}$, we use the value of $H_{c2}$=70(10)~T reported for YBa$_{2}$Cu$_{3}$O$_{6.86}$, and the transition sequence is:
\begin{equation}
\mu_{0}H_{\rm{c1}} < \textrm{TP} < 0.034(5) < \textrm{TP}+\textrm{B} < 0.077(11) < \textrm{A}. \nonumber
\end{equation}
This shows both VL structure transitions to occur at a higher fraction of $H_{c2}$ in the more underdoped compound, even though the absolute transition fields are similar. We also point out that while our experiments made use of the highest magnetic fields presently available for SANS, these fields extend only up to limited fractions of $H_{c2}$; for YBa$_{2}$Cu$_{3}$O$_{7}$ the highest applied field of 16.7~T corresponds to $\sim$0.11~$H_{c2}$,~\cite{Cam14} while for YBa$_{2}$Cu$_{3}$O$_{6.85}$ the highest applied field of 16~T corresponds to $\sim$0.23~$H_{c2}$. From this viewpoint, while the observations in the two compounds are comparable to one another on an absolute field scale, the measurements on the more underdoped compound cover a larger portion of the superconducting phase diagram.

Next we discuss the physical origin of the VL structure transitions seen in both compounds for $\mu_{0}H\parallel$~\textbf{c}. The vast majority of theoretical work on VL structure transitions has been developed on model systems with $\mu_{0}H$ perpendicular to a fourfold symmetry plane, since this provides the simplest system in which to treat the expected effects of either (or both) of a Fermi surface anisotropy and $d$-wave nodal gap symmetry.~\cite{Ber95,Xu96,Aff97,Fra97,Kog97a,Shi99,Ich99,Nak02,Hia08,Suz10} For $\mu_{0}H$ perpendicular to a twofold symmetry plane as appropriate for orthorhombic YBa$_{2}$Cu$_{3}$O$_{7-\delta}$, the available theory is limited~\cite{Kog97a,Hir10} and, as far as we are aware, completely non-existent for the case where $\mu_{0}H$ is perpendicular to a twofold plane that includes a nodal gap anisotropy. We further mention that, since field-driven and still unexplained VL transitions have also been observed in other unconventional superconductors for fields applied perpendicular to a twofold symmetry plane,~\cite{Esk97b,Das12} understanding the observed VL transitions for such a symmetry setting assumes an added importance from the more general perspective. In what follows we interpret the observations in YBa$_{2}$Cu$_{3}$O$_{7-\delta}$ in terms of the expectations of available theory.

As mentioned previously, local London theory provides a useful starting point for understanding the coupling between the VL and the crystal anisotropy in high-$\kappa$ superconductors.~\cite{Kog81,Cam88,Thi89} In this approach, an effective mass anisotropy is incorporated into the free energy via a second rank tensor $m_{ij}$. The inclusion of $m_{ij}$ can lead to distortions of isotropic hexagonal VL structures at no energy cost, and thus local theory can provide an explanation for the distortions of the hexagonal VL domains that we observe at low field in YBa$_{2}$Cu$_{3}$O$_{6.85}$. However, local theory has well-noted shortcomings since no preferred VL orientation is expected for $\mu_{0}H\parallel$~\textbf{c}, and consequently no $\mu_{0}H$-induced transitions in the VL coordination can be expected. Since in all real materials the VL displays a preferred orientation at all fields, it is necessary to consider higher-order effects beyond local theory, i.e. nonlocal effects, to explain the observations.

The inclusion of nonlocal corrections in VL theory provides a means for the host system anisotropies, such as the Fermi velocity and superconducting gap, to couple to the VL properties. Consequently, nonlocal theory can be used to explain both observed preferred VL orientations, and $\mu_{0}H$- and temperature-driven VL structure transitions. Extensive theoretical work has been conducted exploring the role of nonlocality on the VL structure that involves varying degrees of approximation. Common approaches have involved developing nonlocal corrections or higher-order gradients into analytic phenomenological London and Ginzburg-Landau models, respectively.~\cite{Ber95,Xu96,Aff97,Fra97,Kog97a,Shi99,Hia08} The most desirable theoretical approach for calculating the VL properties are first-principle calculations within the framework of quasiclassical Eilenberger theory.~\cite{Eil68,Ich99,Nak02,Suz10} In this approach nonlocality is catered for by default, and the Fermi surface and superconducting gap anisotropies can be treated in equal measure. Furthermore, calculations can be done that are valid at any field or temperature. From such calculations it is well-established that for $\mu_{0}H$ perpendicular to a four-fold symmetry plane, anisotropies arising from both the Fermi velocity and a nodal superconducting gap can determine both the VL orientation, and explain $\mu_{0}H$-driven VL structure transitions.~\cite{Ich99,Nak02,Suz10}

In detail, the more recent quasiclassical Eilenberger calculations reported in Ref.~\cite{Suz10} compared the expected $\mu_{0}H$-dependence of the VL structure for cases where \emph{only} a four-fold Fermi surface anisotropy, or \emph{only} a $d$-wave gap anisotropy influences the VL structure. The calculations show that a `generic' low $\mu_{0}H$-dependent behavior of the VL structure can be expected regardless of the origin of the anisotropy. Briefly, the expectation is that with increasing field a first-order transition between a low field hexagonal VL and a 45$^{\circ}$ rotated hexagonal VL occurs at intermediate field. Further increase of the field causes the intermediate field VL to distort smoothly before a second-order lock-in transition occurs into a square VL. The orientation of the square VL is determined by the symmetry of the anisotropy included in the model, with nearest neighbours aligned along the direction of the Fermi velocity minima, or the nodal direction of the gap. This generic expectation has good support from experiments, and can describe the evolution of the low field VL in V$_{3}$Si,~\cite{Yet05} borocarbides (Er,Y,Lu)Ni$_2$B$_2$C systems,~\cite{Esk97,Pau98,Lev02,Dew05} and heavy-fermion CeCoIn$_{5}$.~\cite{Esk03,Bia08}

\begin{figure}
\centering
\includegraphics[width=0.48\textwidth]{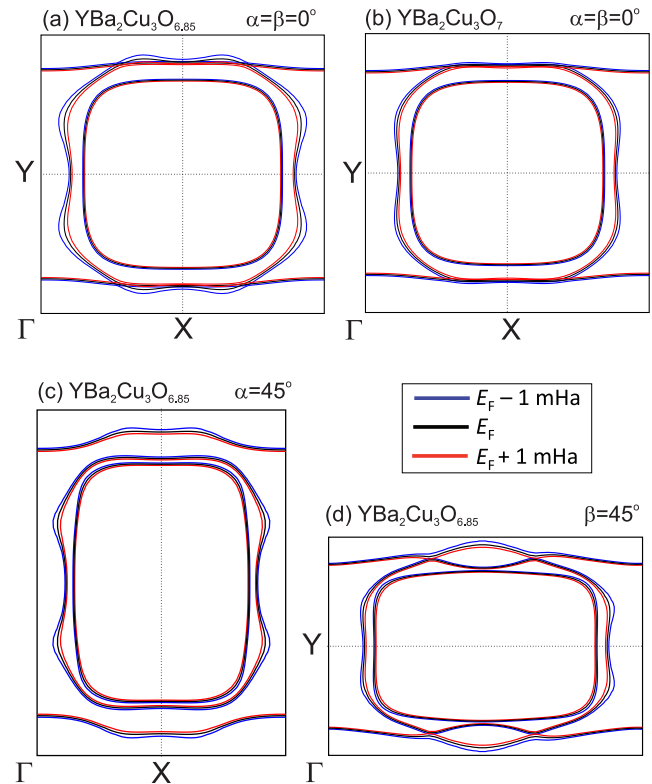}
\caption{(Color) In-plane Fermi surfaces of (a) YBa$_{2}$Cu$_{3}$O$_{6.85}$ and (b) YBa$_{2}$Cu$_{3}$O$_{7}$ obtained from standard LDA calculations using the DMol$^{3}$ code.~\cite{Per92,Del90,Del00} The standard settings included DND numerical local orbital basis sets, and a 4 4 2 $\Gamma$ centered $k$-points mesh. The Fermi surface contours were evaluated on a 50 x 50 mesh in the quarter plane. The calculations for the O$_{6.85}$ stoichiometry was simulated in the YBa$_{2}$Cu$_{3}$O$_{7}$ cell, but with a partial band-filling corresponding to 0.3 holes per unit cell. (c) and (d) show the calculated Fermi surface cuts in the plane perpendicular to $\mu_{0}H$ for YBa$_{2}$Cu$_{3}$O$_{6.85}$, and for the cases of $\alpha$=45$^{\circ}$ and $\beta$=45$^{\circ}$ respectively. In all panels, direction $\Gamma$-X is parallel to the crystal \textbf{a}$^{\ast}$-axis, while $\Gamma$-Y is parallel to the crystal \textbf{b}$^{\ast}$-axis. The black lines denote Fermi energy contours, while the blue (red) lines denote energy contours at $E_{\rm{F}}$-1~mHa ($E_{\rm{F}}$+1~mHa), where 1~mHa is 0.001~Hartree (Ha) units.}
\label{Fig:10}
\end{figure}

From the above discussion, it is clear that when the directions of the Fermi velocity minima and gap nodes are close or coincide, attributing either source of anisotropy as controlling the VL orientation is challenging; just such a situation arises for the case of two-fold symmetric YBa$_{2}$Cu$_{3}$O$_{6.85}$ and YBa$_{2}$Cu$_{3}$O$_{7}$. In Fig.~\ref{Fig:10} we show the Fermi surfaces of each compound calculated within the local-density approximation (LDA) that is strictly valid only for zero-field. Each Fermi surface is dominated by large quasi-cylindrical bonding and antibonding bands arising from the CuO$_{2}$ planes, and a quasi one-dimensional CuO chain band dispersing along $\Gamma-$X that touches the anti-bonding band. The anisotropy of the dominant bonding and antibonding bands in the Brillouin zone is expected to yield Fermi velocity minima along $\langle 110\rangle$ directions. However, a contribution due to superconducting chain states will shift the zone-averaged Fermi velocity minima towards $\langle 100\rangle$. Concerning the gap symmetry, both phase-sensitive~\cite{Kir06} and tunnelling~\cite{Smi05} measurements done on YBa$_{2}$Cu$_{3}$O$_{7-\delta}$ show that the zero-field superconducting gap is of $d_{x^{2}-{y}^{2}}+s$ symmetry. The $d_{x^{2}-{y}^{2}}$ component with nodes along $\langle 110\rangle$ dominates, yet the finite $s$-wave component induced by the crystal orthorhombicity causes an observed shift in the nodal positions also towards $\langle 100\rangle$. Therefore, by itself each of Fermi surface and gap anisotropy are expected to dictate a similar field-dependence of the VL structure in YBa$_{2}$Cu$_{3}$O$_{7-\delta}$.

Bearing in mind the expected similar influences of the Fermi surface and gap anisotropies on the $\mu_{0}H$-dependent VL structure, it is difficult to reconcile the observation of \emph{successive} first-order reorientation transitions like those observed in YBa$_{2}$Cu$_{3}$O$_{7-\delta}$ with the expectation of any theory. As mentioned above, according to the calculations presented in Ref.~\cite{Suz10} each type of anisotropy by itself could cause the expected \emph{generic} field-evolution of the VL structure that displays a \emph{single} low field first-order reorientation transition, and a higher field second-order lock-in transition. In the case of a two-fold symmetry in the Fermi velocity, a similar generic sequence was found from a nonlocal London theory,~\cite{Kog97a} although it was shown later that the high-field continuous transition expected within the model must become forbidden on grounds of symmetry and in reality likely becomes replaced by a smooth crossover.~\cite{Whi11} Indeed, as shown in Fig.~2 for YBa$_{2}$Cu$_{3}$O$_{6.85}$, the field-dependence of the opening angle for structure A, $\phi$ varies smoothly over the high-field range, and above 90$^{\circ}$ tends toward a saturation orientation at the very highest fields.

On considering the above discussion, here we propose that the high field first-order transition seen in YBa$_{2}$Cu$_{3}$O$_{7-\delta}$ corresponds to the ``generic'' first-order transition expected in theory that can in principle be driven by either the Fermi surface or gap anisotropy.~\cite{Suz10} In contrast, we suggest that the low field transition arises instead as a consequence of a $\mu_{0}H$-dependence of the in-plane electronic structure particular to YBa$_{2}$Cu$_{3}$O$_{7-\delta}$, namely the superconducting DOS associated with the CuO chains.

Theoretical work aimed at exploring the role of the chains in the vortex state of YBa$_{2}$Cu$_{3}$O$_{7-\delta}$ included calculations of the magnetic-field dependence of the coherence length, and hence vortex core size, in a bilayer model composed of plane and chain layers.~\cite{Atk08} The calculations revealed that at very low field, the vortex core shape in the chain layer is orthorhombically distorted so that it is two times longer along the chain direction (\textbf{b}-axis) than the \textbf{a}-axis. On the other hand, there is little orthorhombicity of the vortex core shape in the CuO$_{2}$ plane. Nonetheless, the net core shape is significantly longer along the \textbf{b}-axis than the \textbf{a}-axis. By increasing the field, both the core size in the chains, and hence the net core anisotropy, are expected to reduce with increasing field up to a critical field denoted $B^{\ast}$. The authors estimated $B^{\ast}\sim1.5$~T as denoting a critical field above which the vortex cores in the chains overlap. Physically this amounts to the quenching of a component of a small superconducting gap on the chains, and hence a modification in the superconducting DOS. We note that $B^{\ast}$ is in reasonable agreement with the low VL structure transition field seen in both YBa$_{2}$Cu$_{3}$O$_{6.85}$ and YBa$_{2}$Cu$_{3}$O$_{7}$, leading us to expect that this transition is driven by a $\mu_{0}H$-dependent suppression of the core anisotropy. Verification of this expectation requires careful numerical calculations that make use of realistic material parameters for YBa$_{2}$Cu$_{3}$O$_{7-\delta}$. In particular, calculating the spatial distribution of the local DOS near the Fermi energy, and its $\mu_{0}H$-dependence, are key for exploring the stability of the possible VL structures. On the experimental side, verification could be obtained from a direct measurement of the low-field-dependent vortex core anisotropy by using scanning tunnelling microscopy.~\cite{Mag95,Sud14}

The proposal that the low field transition is driven by a $\mu_{0}H$-dependent core anisotropy can be shown as consistent with the SANS measurements at finite $\alpha$/$\beta$. From Fig.~\ref{Fig:8} we see the tendency for the low field transition to shift to lower fields for increasing $\alpha$, and higher fields for increasing $\beta$. Qualitatively this behavior can be understood by considering how the angle evolution of the Fermi surface morphology affects the effective core anisotropy in the plane perpendicular to the applied field. To put this on a firmer footing, Figs.~\ref{Fig:10}(c)-(d) show further LDA calculations of the Fermi surface cuts in planes perpendicular to the applied field for $\alpha$ and $\beta$=45$^{\circ}$. As expected, at finite $\alpha$ and $\beta$ the bonding and antibonding bands become elongated along the direction perpendicular to the rotation compared with the case $\alpha=\beta=0$. On the $\alpha$ side, the increase in the effective mass along the direction including the chain Fermi surface will lead to a reduction in the net coherence length along this direction, and hence result in a reduction of the core anisotropy compared with the case of $\alpha=\beta=0$. Consequently, in accordance with the expectation of Ref.~\cite{Atk08}, a reduced core anisotropy favors the observed reduction in the transition field for finite $\alpha$. On the $\beta$ side, the converse effect is observed because the effective mass increases in the direction \emph{perpendicular} to the chains. This can then be expected to lead to the observed enhancement of the low field-stability of structure A.

The angle-dependent measurements also shed light on the underlying nature of the high field transition for $\mu_{0}H\parallel$~\textbf{c}. Were the high-field transition driven solely by a nodal gap anisotropy, increasing $\alpha$ and $\beta$ would be expected to have a largely similar effect on the transition field since each rotation plane includes an anti-nodal direction. In particular, by rotating the in-plane gap anisotropy away from the field direction it can be expected that a higher magnetic field, and so shorter intervortex distance, is required for the relevant nonlocal interaction to drive the structure transition. Since like the low transition field, the high transition field shifts in opposite directions for the two rotation directions, this confirms that the Fermi surface morphology plays a key role in the high-field transition in YBa$_{2}$Cu$_{3}$O$_{7-\delta}$. It also shows that a significant twofold electronic anisotropy in the superconducting DOS, likely arising from persistent CuO chain superconductivity,~\cite{Atk95,Xia96,Atk99,Atk08} survives until the high-field region.

Having established that the Fermi surface morphology influences strongly the observed VL structure transitions in YBa$_{2}$Cu$_{3}$O$_{7-\delta}$ for $\mu_{0}H\parallel$~\textbf{c}, we finally discuss the role of the nodal gap anisotropy. This anisotropy is generally expected to become increasingly important at higher fields due to the increasing prevalence with vortex density of the vortex lattice effect (VLE).~\cite{Ich99,Suz10} This physical mechanism, which describes the tunnelling of quasiparticles between vortices, is a peculiarity of the intervortex interaction in a superconductor with an in-plane nodal gap, and is most efficient when vortex nearest neighbours are aligned along the nodal directions. In the theory of a fourfold symmetric and purely $d$-wave superconductor, the VLE is expected to lead to the stabilization of a perfectly square VL structure for fields above $\sim$0.15$H_{c2}$.~\cite{Ich99} Note that this expected transition field as a fraction of $H_{c2}$ is significantly higher than the transition fields observed by SANS in either YBa$_{2}$Cu$_{3}$O$_{6.85}$ or YBa$_{2}$Cu$_{3}$O$_{7}$. This therefore brings into question any involvement of the nodal gap anisotropy in driving the high field transition.

Nonetheless, since in our SANS experiments on YBa$_{2}$Cu$_{3}$O$_{6.85}$ we have explored the VL for applied fields up to $\sim$0.23~$H_{c2}$, we can expect that the observed VL properties in the high field range are indeed sensitive to the gap symmetry, even if it is does not drive the high field transition. From this viewpoint, it is therefore interesting to note that, as in YBa$_{2}$Cu$_{3}$O$_{7}$,~\cite{Cam14} at the highest fields the opening angle for structure A extends beyond the 90$^{\circ}$ value expected for a perfectly square coordination. If the very high field VL coordination is indeed determined by the nodal gap anisotropy, then the nodal directions at high field can be concluded to have shifted to the opposite sides of the $\langle110\rangle$ directions compared with the zero-field case.~\cite{Kir06,Smi05} Full clarification of this intriguing high field behavior of the VL structure in YBa$_{2}$Cu$_{3}$O$_{7-\delta}$ would certainly benefit from both new theoretical work and high field experiments.

\section{Summary}
\label{sec:5Summ}
In summary, using the small-angle neutron scattering (SANS) technique we have studied the magnetic field-dependence at 2~K of the VL structure in a sample of detwinned and lightly underdoped YBa$_{2}$Cu$_{3}$O$_{6.85}$. For magnetic fields applied parallel to the crystal \textbf{c}-axis ($\mu_{0}H\parallel$~\textbf{c}), measurements were conducted for fields up to 16~T, and we observed two field-driven and first-order VL structure transitions at 2.37(2)~T and 5.37(2)~T. These transitions are analogous to those reported in detwinned and lightly overdoped YBa$_{2}$Cu$_{3}$O$_{7}$ at 2.3(2)~T and 6.7(2)~T, and it can be expected that the physics behind the transitions in the two compounds is largely similar. In YBa$_{2}$Cu$_{3}$O$_{6.85}$ further measurements at 2~K, and up to 10~T in applied field were done after rotating the direction of applied field away from the \textbf{c}-axis. Dependent on whether the field is rotated around the orthogonal \textbf{a}$^{\ast}$- or \textbf{b}$^{\ast}$-axes, we observed a strong and opposing angle-dependence of the two transition fields, and consequently markedly different VL structure phase diagrams as functions of both field and rotation angle.

From our observations, and in conjunction with available theory, we have argued here that the low-field VL structure transition in both YBa$_{2}$Cu$_{3}$O$_{6.85}$ and YBa$_{2}$Cu$_{3}$O$_{7}$ arises as a consequence of a strong low field-dependence of a vortex core anisotropy that arises from the CuO chain superconducting density of states. Due to the proposed involvement of the chain states, this low field transition can be considered as particular to YBa$_{2}$Cu$_{3}$O$_{7-\delta}$.

In contrast, and once the pronounced low-field core anisotropy is suppressed on increasing the field, we propose that the high-field VL structure transition in both compounds more likely corresponds to the `generic' first-order VL reorientation transition expected in theory to be driven either by an anisotropy in the Fermi surface or a nodal gap. Although we are unable to unambiguously assign a particular anisotropy as driving the high field transition, the contrasting angle-dependences of the transition fields on the direction of field-rotation away from the \textbf{c}-axis allow us to demonstrate that this transition is indeed dependent on the Fermi surface morphology. By again comparing our experimental results to those of theory, we expect that the clearest evidence for an influence of the nodal gap structure on the VL properties is most likely away from the second transition field, and instead more in the range of the highest accessible fields. This can be checked by further field-angle SANS VL measurements done for very high fields above 10~T.

Finally, and perhaps most strikingly, on the approach to the highest explored fields in YBa$_{2}$Cu$_{3}$O$_{6.85}$ with $\mu_{0}H\parallel$~\textbf{c}=16~T, the precise field-evolution of the VL coordination is observed to tend towards a saturation orientation. This very high field VL coordination is not consistent with being stabilized by either the established superconducting gap or Fermi surface anisotropies at zero-field. This ultimately demonstrates the strong magnetic field-dependence of the superconducting density of states in YBa$_{2}$Cu$_{3}$O$_{7-\delta}$ that is reflected by the VL structure phase diagram. Overall it can be stated that the accumulated SANS observations on YBa$_{2}$Cu$_{3}$O$_{7-\delta}$ illustrate the richness of the VLs in even `prototypical' unconventional superconductors, but which still call for further experimental and theoretical studies aimed at clarifying the mechanisms that dictate the novel field-dependent properties.

\section{Acknowledgements}
SANS experiments were performed at the Institut Laue-Langvein, Grenoble, France, the Swiss spallation neutron source (SINQ), Paul Scherrer Institut (PSI), Switzerland, and the NG3-SANS instrument at the NIST Center for Neutron Research, Gaithersburg, USA. We thank S.~Pheiffer, P.~Butler, D.~Dender, and R.~Dimeo for support at NIST, and especially A.~Jackson for providing a hand with the experiments there. J.-T.~Park is acknowledged for assistance with the sample preparation at MPI Stuttgart. We further thank J.~Kohlbrecher and Ch.~R\"{u}egg for support at PSI, and K.~Machida and I.~Maggio-Aprile for discussions. We acknowledge financial support from the EPSRC of the UK, the University of Birmingham, the Swiss NCCR and its program MaNEP, the Swiss National Science Foundation, and from the European Commission under the 6$^{th}$ Framework Programme though the Key Action: Strengthening the European Research Area, Research Infrastructures, Contract No. RII3-CT-2003-505925. This work constitutes part of the PhD thesis of NGL.

\end{document}